\documentclass[twocolumn,aps,showpacs,floatfix,prc]{revtex4}
\usepackage[dvips]{epsfig}
\newcommand{\etal}{{\it et al.}}
\begin{document}
 
\title{$J/\psi$ suppression and $p_T$ spectra in RHIC and LHC energy collisions}
\author{A. K. Chaudhuri}
\email[E-mail:]{akc@veccal.ernet.in}
\affiliation{Variable Energy Cyclotron Centre,\\ 1/AF, Bidhan Nagar,
Kolkata 700~064, India}

\begin{abstract}
In a hydrodynamic model, we have studied $J/\psi$ production in Au+Au/Cu+Cu collisions at RHIC energy $\sqrt{s}$=200 GeV. At the initial time, $J/\psi$'s are randomly distributed in the fluid.  As the fluid evolve in time, the free streaming $J/\psi$'s are dissolved if the local fluid  temperature exceeds a threshold temperature $T_{J/\psi}$. Sequential melting of charmonium states
($\chi_c$, $\psi\prime$ and $J/\psi$), with melting temperatures $T_{\chi_c}=T_{\psi\prime} \approx 1.2T_c$, $T_{J/\psi} \approx2T_c$ and  feed-down fraction $F\approx 0.3$,  explains the 
PHENIX data  on the centrality dependence of  $J/\psi$ suppression in   Au+Au collisions. $J/\psi$ $p_T$ spectra and the nuclear modification factor in  Au+Au collisions are also well explained in the model.   The model however over predict  centrality dependence of $J/\psi$ suppression in Cu+Cu collisions by 20-30\%. The
$J/\psi$ $p_T$ spectra are under predicted by 20-30\%.  The model predict that
in central Pb+Pb collisions at LHC energy, $\sqrt{s}$=5500 GeV, $J/\psi$'s are suppressed by a factor of $\sim$ 10. 
The model predicted $J/\psi$ $p_T$ distribution in Pb+Pb collisions at LHC is similar to
that in Au+Au collisions at RHIC.
 \end{abstract}  

\maketitle
 
\section{introduction} 
 
High energy heavy ion collisions give the opportunity to study QCD matter at high density and temperature. Recent experiments at Relativistic Heavy Ion Collider (RHIC) at Brook Haven National Laboratory, give strong indications that in Au+Au collisions, a collective QCD matter is formed \cite{BRAHMSwhitepaper,PHOBOSwhitepaper,PHENIXwhitepaper,STARwhitepaper}. However, whether the matter can be qualified as the lattice QCD predicted Quark-Gluon Plasma \cite{lattice} is still a matter of debate. $J/\psi$ suppression is long
recognized as an important tool  to  identify  the  possible
phase transition to quark-gluon plasma.   Matsui and Satz  \cite{Matsui:1986dk}, 
predicted that in presence of quark-gluon plasma  (QGP),  binding
of a $c\bar{c}$  pair  into  a  $J/\psi$  meson will be hindered,
leading to the  so  called  $J/\psi$  suppression  in  heavy  ion
collisions  \cite{Matsui:1986dk}. It was later realised that $J/\psi$'s are not dissolved just at the critical temperature $T_c $.   Lattice based potential models indicate that dissociation
temperature for $J/\psi$ is $T_{J/\psi}\approx$ 2.1 $T_c$ \cite{Satz:2006kb}. The excited states
$\chi_c$ and $\psi\prime$ on the other hand are dissolved close to the critical temperature $T_{\chi_c} \approx 1.2T_c$, $T_{\psi\prime} \approx 1.1T_c$ \cite{Satz:2006kb}.

PHENIX  collaboration have measured $J/\psi$ yield in p+p \cite{Adler:2003qs,Adler:2005ph,Adare:2006kf}, d+Au \cite{Adler:2005ph,Adare:2007gn}, Au+Au \cite{Adler:2003rc,Adare:2006ns} and Cu+Cu \cite{Adare:2008sh} collisions at RHIC. $J/\psi$'s are suppressed in d+Au,
Au+Au and Cu+Cu  collisions. Suppression in d+Au collisions is due to a number of cold nuclear matter (CNM) effects, e.g. gluon shadowing, gluon energy loss,
$J/\psi$-nucleon inelastic collision. $J/\psi$ suppression in hot nuclear matter
is measured in Au+Au and Cu+Cu collisions.  In Au+Au/Cu+Cu collisions,
suppression increases with centrality of the collisions and in very peripheral collisions, suppression is consistent with CNM effects. Also  $J/\psi$'s are more suppressed in forward rapidity than at mid rapidity. Measurements also indicate
that models \cite{ Thews:2000rj,Braun-Munzinger:2000px}, which predict for $J/\psi$ enhancement in RHIC energy collisions,
are not consistent with experiments. On the other hand, phenomenological threshold model \cite{Blaizot:2000ev,Blaizot:1996nq}, which mimic the melting of $J/\psi$'s in a QGP medium, successfully explain the  suppression in Au+Au collisions but not in Cu+Cu collisions \cite{Chaudhuri:2006fe,Chaudhuri:2007qz,Chaudhuri:2007yy}. The threshold model
do not account for the possible recombination of $c\bar{c}$ pairs. 
In \cite{Capella:2007jv} PHENIX data were analysed in the comover model. It was
claimed that the comover model
, including recombination, is consistent with the centrality dependence of
$J/\psi$ suppression both in Au+Au and Cu+Cu collisions \cite{Capella:2007jv}. However, we noticed that
in mid-rapidity Au+Au collisions, the model do not reproduce the sudden increase of suppression beyond $N_{part} \sim$ 150.  

Recently, following  Gunji \etal \cite{Gunji:2007uy}, we have developed
   a 'Hydro+$J/\psi$' model to simulate $J/\psi$ production in nuclear collisions \cite{Chaudhuri:2008qq}. The QGP fluid evolves in 2+1 dimensions. At the initial time,   $J/\psi$'s are randomly distributed in the transverse plane. As the fluid evolve, free streaming $J/\psi$'s are melted if the local fluid temperature exceeds a critical value. A considerable percentage of observed $J/\psi$'s are from decay of higher states, $\chi$ and $\psi\prime$   \cite{Abt:2002vq}. 
The experimental $J/\psi$ suppression pattern in mid-rapidity Au+Au collisions is well explained by sequential melting of $\chi_c$, $\psi\prime$ and $J/\psi$ in the dynamically expanding fluid. 
The estimated melting temperatures, $T_{J/\psi}=2.1 T_c$,
$T_{\chi_c}=T_{\psi\prime}=1.1 T_c$   are in agreement with the lattice motivated calculations \cite{Satz:2006kb}. The fraction of the higher states $(\chi_c+\psi^\prime)$ is estimated to be $F=0.3$. The model however fails to reproduce the centrality dependence of $J/\psi$ suppression in Cu+Cu collisions.
Suppression is over predicted.
In the present paper, we have improved upon the model and studied   transverse momentum distribution of $J/\psi$'s in Au+Au  and in Cu+Cu collisions.   
The improved model, in addition to the centrality dependence of $J/\psi$ suppression, well reproduces the $J/\psi$ $p_T$ spectra in Au+Au collisions. However,  in Cu+Cu collisions, the improved model over predict $J/\psi$
suppression. The $p_T$ spectra is under predicted. We have also used the model to
predict for the centrality dependence of $J/\psi$ suppression and $p_T$ spectra    in Pb+Pb collisions at LHC energy.   In the model, $J/\psi$'s are greatly suppressed in LHC energy.  

The plan of the paper is as follows: in section \ref{sec2}, we briefly describe the 'Hydro+$J/\psi$' model.   In section \ref{sec3}, we have analysed  PHENIX data on the centrality dependence of $J/\psi$ suppression, $J/\psi$ $p_T$ spectra in Au+Au and Cu+Cu collisions.
In section \ref{sec4} we give predictions for centrality dependence of $J/\psi$ suppression in Pb+Pb collisions at LHC. Predictions for $J/\psi$ $p_T$ spectra, in 0-20\%, 20-40\%, 40-60\% and 60-92\%  Pb+Pb collisions at LHC
energy are also given in section \ref{sec4}. Lastly.
summary and conclusions are drawn in section \ref{sec5}.
 
\section{The hydrodynamic model for $J/\psi$ suppression} \label{sec2}

Details of the hydrodynamic model used here can be 
found in \cite{QGP3,Chaudhuri:2008qq}. Briefly,
it is assumed that 
in high energy nuclear collisions, a deconfined phase (QGP) is produced, which expands, cools, undergoes 1st order phase transition to hadronic fluid at the critical temperature ($T_c$=164 MeV) and then further cools to freeze-out at temperature $T_F$=130 MeV.  
Assuming longitudinal boost-invariance, the space-time evolution of the fluid is obtained by solving the energy-momentum conservation equation $\partial_\mu T^{\mu\nu}=0$ \cite{QGP3}. Assumption of boost-invariance limits the model to mid-rapidity region only. $J/\psi$'s are assumed to be produced in initial NN collisions. They are free streaming, unless dissolved in the QGP medium.
Hydrodynamic models require   energy density, fluid velocity distributions at the initial time $\tau_i$. The initial energy density of the fluid in  the transverse plane was parameterised as \cite{QGP3,Chaudhuri:2008qq}, 

\begin{equation} \label{eq1}
\varepsilon({\bf b},x,y) =\varepsilon_0 [0.75 N_{part}({\bf b},x,y) +.25 N_{coll}({\bf b},x,y)]
\end{equation}

\noindent where $N_{part}({\bf b},x,y)$ and $N_{coll}({\bf b},x,y)$ are the transverse profile for the participant number and binary collisions number in an  impact parameter {\bf b}  Au+Au collision. The initial fluid velocity was assumed to be zero, $v_x({\bf b},x,y)=v_y({\bf b},x,y)=0$. The constant $\varepsilon_0$ depend only on the collision energy and not on centrality of the collisions. The initial time $\tau_i$ and the constant $\varepsilon_0$ are chosen to reproduce the $p_T$ distribution of identified particles in central Au+Au collisions. 
 For b=0 Au+Au collisions, it correspond to central energy density, $\varepsilon$=30 $GeV/fm^{3}$ or central entropy density  $S_{ini}$=110 $fm^{-3}$, at the initial time $\tau_i$=0.6 fm/c. Solution of hydrodynamic equations require an equation of state. We have used the equation of state EOS-Q \cite{QGP3},   incorporating 1st order phase transition with critical temperature $T_c$=164 MeV. 

To obtain the survival probability of $J/\psi$'s in
an expanding medium, we proceed as follows:
at the initial time $\tau_i=0.6 fm/c$,    we  randomly distribute a fixed number of $J/\psi$'s in the transverse plane. They are assumed to be free streaming \cite{note1} unless dissolved in the medium. 
Each $J/\psi$ is characterised by 4 random numbers. Two random numbers ($R_1,R_2$) indicate its transverse position (${\bf r}_\perp$),
and two random numbers ($R_3,R_4$) its transverse momentum $\vec{p}_T$. In     \cite{Gunji:2007uy,Chaudhuri:2008qq}, $J/\psi$'s are spatially distributed according to initial fluid temperature profile. However, initial NN collisions produces $c\bar{c}$ pairs, which later evolve to $J/\psi$. The formation time of $J/\psi\sim$ 0.5 fm, incidentally is similar to the thermalisation time of the QGP fluid. Then, spatial distribution of initial $J/\psi$'s may not follow the spatial distribution of the fluid temperature,
rather   it is more likely follow the transverse profile of the binary collision number.
Presently, we distribute the random numbers $R_1$ and $R_2$ according to the transverse profile of the number of binary collisions ($N_{coll}$).

PHENIX collaboration have measured $J/\psi$ $p_T$ spectra in p+p collisions at RHIC \cite{Adler:2003qs,Adler:2005ph,Adare:2006kf}. The power law \cite{Yoh:1978id},

\begin{equation}\label{eq2}
B\frac{d\sigma}{dyd^2p_T}=\frac{A}{[1+(p_T/B)^2]^6} (nb/GeV^2),
\end{equation}

\noindent with $A=4.23$ and $B=4.1$ rather well describe the invariant distribution of measured $J/\psi$'s in p+p collisions . The random number $R_3$ is distributed following the power law Eq.\ref{eq2}. The random number $R_4$ is 
distributed uniformly within [0-2$\pi$].  
 
The survival probability of a $J/\psi$ inside the expanding QGP is calculated as \cite{Gunji:2007uy},

\begin{equation} \label{eq3}
S_{J/\psi}(\tau)=exp\left[-\int_{\tau_i}^\tau \Gamma_{dis}(T({\bf r}_{\perp}(\tau^\prime))) d\tau^\prime \right]
\end{equation}

\noindent where $T({\bf r}_\perp)$ is the temperature of the fluid at the transverse position $r_\perp$, $\Gamma_{dis}(T)$ is the decay width of $J/\psi$ at temperature $T$. $\tau_i$ is the initial time for hydrodynamic evolution. We continue the evolution till the freeze-out temperature $T_F$=130 MeV. 
In our earlier work \cite{Chaudhuri:2008qq},
for the decay width $\Gamma_{dis}$ we had used the simplest prescription,

\begin{eqnarray} \nonumber
\Gamma_{dis}(T) &= & \infty; \hspace{1cm} T>T_{J/\psi} \\
\Gamma_{dis}(T) &= &0         \hspace{1cm}T<T_{J/\psi} \label{eq4}
\end{eqnarray}

The decay width Eq.\ref{eq4}, is rather rough. At the melting temperature $T_{J/\psi}$, it abruptly changes from zero to $\infty$. Gunji et al \cite{Gunji:2007uy}, in their analysis PHENIX Au+Au data used the following form for the decay width,

\begin{eqnarray} \nonumber
\Gamma_{dis}(T) &= & \infty; \hspace{1cm} T>T_{J/\psi} \\
\Gamma_{dis}(T) &= &\alpha (T/T_c -1)^2; \hspace{1cm}T<T_{J/\psi} \label{eq5}
\end{eqnarray}

In Eq.\ref{eq5}, $\alpha$  is the thermal width of the state at $T/T_c$=2. NLO perturbative calculations suggest that
$\alpha > 0.4 GeV$ \cite{Park:2007zza}. Gunji et al \cite{Gunji:2007uy} observed that for $\alpha \geq 0.4$, the PHENIX data are not well described. Data are best described with $\alpha$=0.1. We note that Eq.\ref{eq5} do not take into account of regeneration of charmonium states. Lattice QCD calculations \cite{Asakawa:2003re}   indicate D- and B-like states can exist in sQGP. D- and B-like states in sQGP can   have direct impact on $J/\psi$ production, facilitating regeneration.
They can also provide for the resonant cross section for heavy quarks and play an 
play an essential part in thermalisation and collective flow of charm and bottom quarks \cite{van Hees:2005wb}. 
 
\begin{figure}[t]
\includegraphics[bb=11 173 575 769
 ,width=0.8\linewidth,clip]{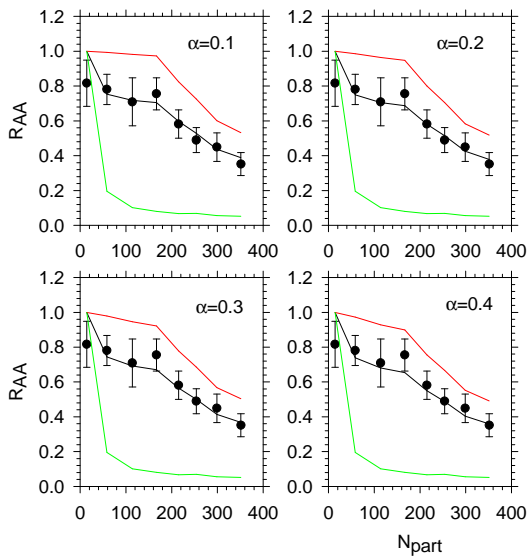}%{auraa.eps}
\caption{(color online) The filled
circles are the PHENIX data   \cite{Adare:2006ns,Leitch:2007wa,Gunji:2007fi} for the centrality dependence of nuclear modification factor ($R_{AA}$) for $J/\psi$ in mid-rapidity Au+Au collisions. 'Hydro+$J/\psi$' model predictions for four values of the decay width controlling parameter $\alpha$=0.1, 0.2, 0.3, and 0.4 are shown in 4 panels. In each panel, the red and green lines are $R_{AA}$ for direct $J/\psi$'s and excited states $\psi\prime$ $\chi_c$. The melting temperatures are $T_{J/\psi}=2T_c$, $T_{\psi\prime}=T_{\chi_c}=1.2T_c$. The black lines are $R_{AA}$ assuming 30\% feed-down.}
\label{F1}
\end{figure}

Any model of $J/\psi$ suppression must account for the experimental observation that a substantial fraction of the measured $J/\psi$'s are from decay of the excited charmonium states $\chi_c$ and $\psi\prime$ \cite{Abt:2002vq}. 
To calculate the survival probability for the excited states $\chi_c$ and $\psi^\prime$, we use the same procedure as described above
for the ground state $J/{\psi}$. 
Above $T_{\chi_c}$ and $T_{\psi\prime}$, the excited states
$\chi_c$ and $\psi^\prime$ are assumed to melt. Further noting that the lattice motivated calculations \cite{Satz:2006kb} indicate $T_{\chi_c} \approx T_{\psi^\prime}$, we define a
common temperature $T_\chi=T_{\chi_c}=T_{\psi\prime}$, above which all the states $\chi_c$ and $\psi^\prime$ are dissolved. For feed-down fraction $F$, the $J/\psi$ survival probability is then obtained as,

\begin{equation}
S_{QGP}=(1-F) S_{J/\psi} + F S_\chi
\end{equation}

As mentioned earlier, PHENIX collaboration in d+Au collisions has studied cold nuclear matter effect on $J/\psi$ suppression \cite{Adler:2005ph,Adare:2007gn}. $J/\psi$'s are suppressed in d+Au collisions also. The suppression is consistent with Glauber model of nuclear absorption with $J/\psi$-nucleon absorption cross-section $\sigma_{abs} =2\pm 1$ mb   \cite{Vogt:2005ia}. If cold nuclear matter effect is taken into account, the survival probability of $J/\psi$ can be obtained as,

\begin{equation} S_{J/\psi}= S_{QGP}\times S_{CNM}, \end{equation}

\noindent where $S_{CNM}$ is the survival probability in cold nuclear matter.

 \begin{figure}[t]
\includegraphics[bb=28 206 580 769
 ,width=0.8\linewidth,clip]{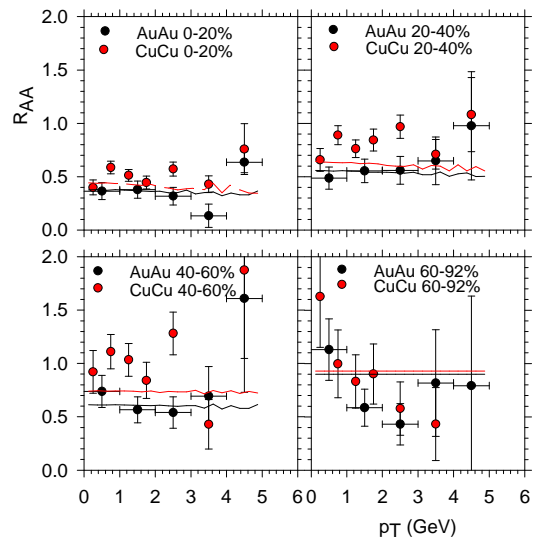} 
\caption{(color online) Black and red circles are PHENIX data for $p_T$ dependence of nuclear modification factor ($R_{AA}$) for $J/\psi$'s in
0-20\%, 20-40\%, 40-60\% and 60-90\% centrality Au+Au \cite{Adare:2006ns} and Cu+Cu  \cite{Adare:2008sh} collisions respectively. Errors are statistical only. The black and red 
lines are the 'Hydro+$J/\psi$' model predictions for Au+Au and Cu+Cu collisions.
Decay width controlling parameter $\alpha$=0.2.  CNM  effects are included (see text).}
\label{F2}
\end{figure}

\section{$J/\psi$ suppression in RHIC energy} \label{sec3}

\subsection{Au+Au collisions}

Experimental survival probability for $J/\psi$ in mid-rapidity Au+Au collisions
are shown in Fig.\ref{F1}, where gluon shadowing and nuclear absorption
with $\sigma_{abs}$=1 mb is taken into account as cold nuclear matter (CNM) effect \cite{Adare:2006kf,Leitch:2007wa,Gunji:2007fi}.  
In Fig.\ref{F1}, we have shown the results of the present 'Hydro+$J/\psi$' model. They are obtained for melting temperatures $T_{J/\psi}=2T_c$, $T_{\psi\prime}=T_{\chi_c}=1.2T_c$.  Results are shown for four values of
the decay width controlling parameter $\alpha$=0.1, 0.2, 0.3 and 0.4. The red and green lines are survival probability for the states $J/\psi$ and $\chi$. The black line is the combined suppression with 30\% feed-down. Sequential melting of charmonium states explains the PHENIX data on the centrality dependence of $J/\psi$ suppression in Au+Au collisions.
As in \cite{Gunji:2007uy}, $J/\psi$ suppression depend on the decay width controlling parameter $\alpha$, increasing with $\alpha$.  However, the dependence
seems to be smaller than found by Gunji et al \cite{Gunji:2007uy}. For example, in the most central collision ($N_{part}=325)$ changing $\alpha$ from 0.1 to 0.4 changes $R_{AA}$ by less than a few percent. Good description of the data for $\alpha$=0.1-0.4, leave little or no room for charmonium regeneration.  

    \begin{figure}[t]
\includegraphics[bb=24 290 522 771
 ,width=0.8\linewidth,clip]{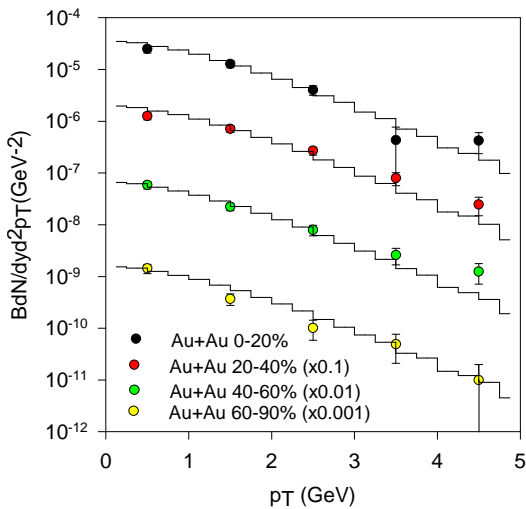} 
\caption{(color online) The colored circles are the PHENIX data \cite{Adare:2006ns} for $J/\psi$ $p_T$ distribution in 0-20\%, 20-40\%, 40-60\% and 60-90\% centrality Au+Au collisions. The black lines are the $J/\psi$ $p_T$ spectra in '$J/\psi$+hydro' model predictions. The CNM effect is included (see text).}
\label{F3}
\end{figure}

PHENIX collaboration have measured $J/\psi$ transverse momentum distribution
    and also the nuclear modification factor ($R_{AA}(p_T)$), 
in  0-20\%, 20-40\% , 40-60\% and 60-92\% centrality Au+Au collisions \cite{Adare:2006ns}.
 In Fig.\ref{F2}, in four panels, PHENIX results for the nuclear modification factor ($R_{AA}$)
in 0-20\%, 20-40\%, 40-60\% and 60-92\% centrality Au+Au collisions are shown. Suppression increases as the collision centrality increases. Also     within the errors, suppression is approximately $p_T$ independent. Note that PHENIX data on $R_{AA}(p_T)$ are not corrected for the CNM effects.   
  We have approximated CNM effect by the Glauber model of nuclear absorption with $\sigma_{abs}$=2 mb \cite{Vogt:2005ia}. In Fig.\ref{F2}, black lines are 'Hydro+$J/\psi$' model prediction. The decay width controlling parameter is $\alpha$=0.2.  Sequential melting of charmonium states correctly reproduces the PHENIX data on the $p_T$ dependence of nuclear modification factor.
Only in very peripheral (60-92\% centrality) collisions, the model do not reproduce the experimental data in $p_T$ range 1-3 GeV. It is not surprising.
Hydrodynamics models are not very accurate at large $p_T$ or in very peripheral collisions \cite{QGP3}. In Fig.\ref{F3}, we have compared model predictions against the PHENIX data on $J/\psi$ $p_T$ distribution in 0-20\%, 20-40\%, 40-60\% and 60-92\% centrality Au+Au collisions. As expected,
the 'Hydro+$J/\psi$' model, with  sequential
melting of charmonium states, also explain the PHENIX data on the $J/\psi$ $p_T$ spectra.

PHENIX collaboration also measured the mean square transverse momentum ($<p_T^2>$) for $J/\psi$.  In table \ref{table1}, PHENIX measurements of $<p_T^2>$  in different centrality ranges of Au+Au collisions are shown. 
Within the errors, $<p^2_T>$ do not show variation with collisions centrality and is approximately same as in pp collisions. As shown in the table, 'Hydro+$J/\psi$' model also do not show any appreciable variation of $<p_T^2>$ with collision centrality. Figs.\ref{F1}, \ref{F2}, \ref{F3} and table \ref{table1}, clearly indicate that sequential melting of charmonium states, with melting temperature $T_{J/\psi}=2T_c$, $T_{\psi\prime}=T_{\chi_c}=1.2T_c$ and feed down fraction F=0.3, explains most of the features of $J/\psi$ production in Au+Au collisions at RHIC energy.

%\begin{widetext}
\begin{table}[h] \label{table1} 
\caption{PHENIX results for $J/\psi$ $<p_T^2>$ are compared with 
'Hydro+$J/\psi$' model calculations. Melting temperatures  of $J/\psi$ and $\chi$ are
$T_{J/\psi}=2T_c$ and $T_\chi=1.2T_c$. The feed down fraction is F=0.3.}
\begin{tabular}{|cccc|}\hline \hline
Percent (\%) & $N_{part}$ & $<p_T^2>_{ex}$ & $<p_T^2>_{th}$   \\
            &             & $(GeV^2)$             & $(GeV^2)$  \\
\hline 
0-20 & 280 & $3.6\pm 0.6\pm 0.1$ & 3.76  \\
20-40 & 140 & $4.6\pm 0.5\pm 0.1$ & 3.80  \\
40-60 & 60  & $4.5\pm 0.7\pm 0.2$ & 3.81  \\
60-92 & 14  & $3.6\pm 0.9\pm 0.2$ & 3.81  \\
p+p    &  2  & $4.1\pm 0.2\pm 0.1$ & 3.87  \\ 
\hline \hline
 \end{tabular} \label{table1}
\end{table}
%\end{widetext} 

\begin{figure}[t]
\includegraphics[bb= 40 286 532 769
 ,width=0.8\linewidth,clip]{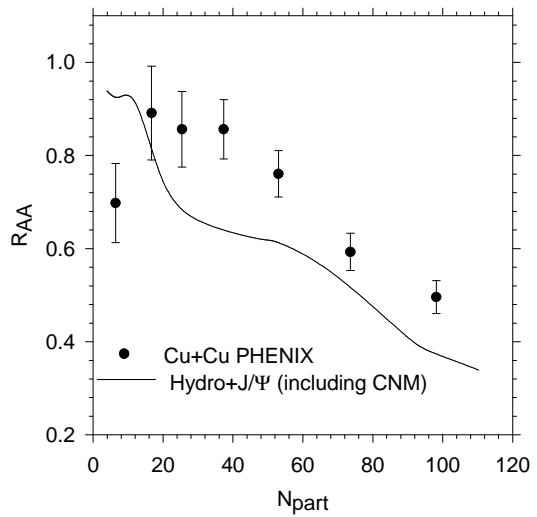}%{curaa.eps}
\caption{The black circles are PHENIX data \cite{Adare:2008sh} on the centrality dependence of $J/\psi$ suppression in Cu+Cu collisions. The black line is the 'Hydro+$J/\psi$' model prediction for the $J/\psi$ suppression in Cu+Cu collisions. }\label{F4}
\end{figure}

\subsection{Cu+Cu collisions}

Recently PHENIX collaboration published their analysis of $J/\psi$ measurements in Cu+Cu collisions \cite{Adare:2008sh}.  
In Fig.\ref{F4}, PHENIX data \cite{Adare:2008sh} on the centrality dependence of $J/\psi$ suppression are shown (the black circles). Data are not corrected for the
cold nuclear matter (CNM) effects.  
In the most central collision, $J/\psi$'s are suppressed by a factor of $\sim$ 2. Survival probability of $J/\psi$ increases as the collision centrality decreases till $N_{part}$=16.7. In more peripheral ($N_{part}$=6.4), collisions, survival probability decreases again. The  decrease in survival probability as
participant number decreases from $N_{part}$=16.7 to 6.4 
is interesting. All the theoretical models predict continuous
increase of suppression as the collision centrality increases. 

 \begin{figure}[t]
\includegraphics[bb=39 291 524 769
 ,width=0.8\linewidth,clip]{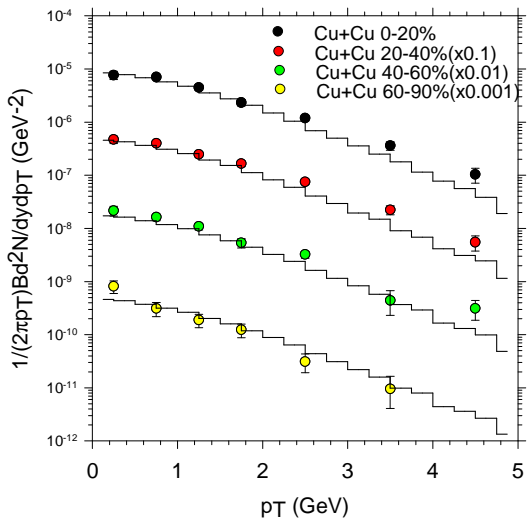}%{cupt.eps}
\caption{ (color online)
The colored symbols are the PHENIX data for $J/\psi$ $p_T$ distribution in 0-20\%, 20-40\%, 40-60\% and 60-90\% centrality Cu+Cu collisions \cite{Adare:2008sh}.  
The black lines are the $J/\psi$  $p_T$ spectra in 'Hydro+$J/\psi$' model.
The melting temperature are $T_{J/\psi}=2.0 T_c$ and $T_\chi=T_{\chi_c}=T_{\psi\prime}=1.2T_c$. The feed-down fraction is $F=0.3$.}  
\label{F5}
\end{figure}

The black line in Fig.\ref{F4} is the 'Hydro+$J/\psi$' model prediction
for the centrality dependence of $J/\psi$ suppression in Cu+Cu collisions. 
The model parameters are same as it was for Au+Au collisions, i.e.
$T_{J/\psi}=2T_c$, $T_{\psi\prime}=T_{\chi_c}=1.2T_c$, $F=0.3$ and $\alpha$=0.1.
We have   approximated the CNM effects by the Glauber model of nuclear absorption with  $J/\psi$-nucleon absorption cross section $\sigma_{abs}$=2 mb.  
Sequential melting of charmonium states
fails to explain the PHENIX data on the centrality dependence of $J/\psi$ suppression. 
In central collisions, the model over predict the suppression by 20-30\%. 
 As discussed earlier, we have neglected charmonium regeneration. If regeneration of $J/\psi$ is important in Cu+Cu collisions, the gap between theory and experiment may be filled by the regenerated charmoniums. However, 
as shown earlier, in Au+Au collisions, there is little or no scope for recombination of $c\bar{c}$ pairs in 'Hydro+$J/\psi$' model.
It
is unlikely that regeneration of charmoniums will be important in a smaller system like Cu+Cu. Indeed, in \cite{Capella:2007jv},
where $J/\psi$ suppression in Au+Au and Cu+Cu collisions were analsyed in the comover model, including recombination, it was observed that recombination effect
reduces suppression by $\sim$50\% in central Au+Au collisions, but reduction is only $\sim$ 10\% in central Cu+Cu collisions.

PHENIX collaboration also measured $J/\psi$ $p_T$ distribution in Cu+Cu collisions \cite{Adare:2008sh}. In Fig.\ref{F5}, we have shown the PHENIX measurements for $J/\psi$ $p_T$ spectra in 0-20\%, 20-40\%, 40-60\% and 60-92\%
centrality Cu+Cu collisions. In Fig.\ref{F5}, the black lines are the $p_T$ spectra in 'Hydro+$J/\psi$' model
with sequential melting of charmonium states. CNM effects are included. PHENIX data are reproduced within 20-30\%. Quality of fit to the data is better judged if nuclear modification factor $R_{AA}(p_T)$ for $J/\psi$'s are compared . In Fig.\ref{F2}, the red symbols are $R_{AA}(p_T)$ for $J/\psi$ in Cu+Cu collisions \cite{note2} and red lines are the model predictions.      Compared to Au+Au collisions, $J/\psi$'s are slightly less suppressed in Cu+Cu collisions. Also, as it is for Au+Au collisions, in Cu+Cu collisions also, $R_{AA}(p_T)$  is approximately $p_T$ independent and increases with centrality .  In 0-20\%, 20-40\% and 40-60\% centrality Cu+Cu collisions, sequential melting of charmonium states over predict the data by 20-30\%.    Evidently, charmonium production mechanism in Cu+Cu collisions is at variance with the charmonium
production in Au+Au collisions.

\begin{figure}[t]
\includegraphics[bb=40 286 532 768
 ,width=0.8\linewidth,clip]{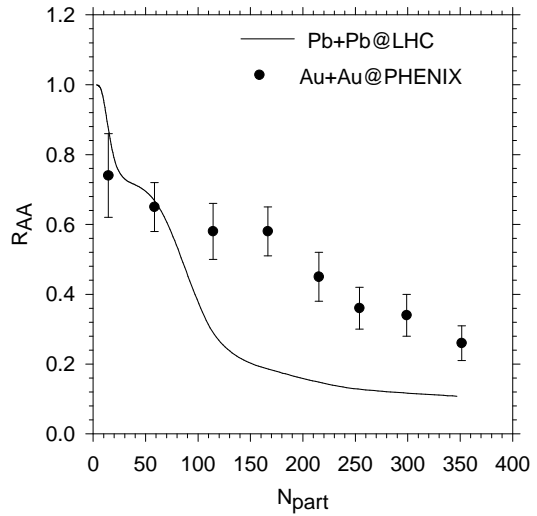}%{pbraa.eps}
\caption{ 'Hydro+$J/\psi$' model predictions for the centrality dependence of $J/\psi$ suppression in Pb+Pb collisions at LHC energy. The solid line is the model prediction with the cold nuclear matter effects.  For comparison,
PHENIX data on the centrality dependence of
 $J/\psi$ suppression in Au+Au collisions \cite{Adare:2006ns,Leitch:2007wa,Gunji:2007fi} are also shown.}
\label{F6}
\end{figure}

\section{$J/\psi$ suppression in Pb+Pb collision at LHC}\label{sec4}

'Hydro+$J/\psi$' model can be used to predict for the $J/\psi$ suppression in Pb+Pb collisions at LHC energy ($\sqrt{s}$=5500 GeV).
The initial conditions for hydrodynamic evolution in Pb+Pb collisions at LHC energy is unknown. We have used the existing knowledge at RHIC energy to extrapolate to
LHC energy and obtained the initial condition for the hydrodynamic evolution.
Central temperature of the fluid in b=0 Pb+Pb collisions is $T_i$=421 MeV (for details see  \cite{Chaudhuri:2008qq}). This can be 
contrasted with central temperature $T_i$=351 MeV in b=0 Au+Au collisions at RHIC. 
The 'Hydro+$J/\psi$' model also require $J/\psi$ $p_T$ distribution in p+p collisions as an input. It is yet to be measured.   While there are many theoretical predictions for $J/\psi$ $p_T$ distribution at LHC energy, most of them are in the framework of pQCD  and predicts for large $p_T$.
 'Hydro+$J/\psi$' model on the other hand is suitable for low $p_T$ particles. Presently, we approximate the $J/\psi$ $p_T$ distribution in p+p collisions at LHC energy by the same power law as it is for RHIC collisions (Eq.\ref{eq2}).
Particle multiplicity increases from RHIC to LHC energy. Assuming logarithmic increase with energy, at LHC energy, $J/\psi$ multiplicity can be increased  by a factor of $\sim$1.6 \cite{Chaudhuri:2008qq}. To be consistent with the increased multiplicity at LHC, we normalise Eq.\ref{eq2} by the factor of 1.6.
 
  \begin{figure}[t]
\includegraphics[bb=32 316 532 795
 ,width=0.8\linewidth,clip]{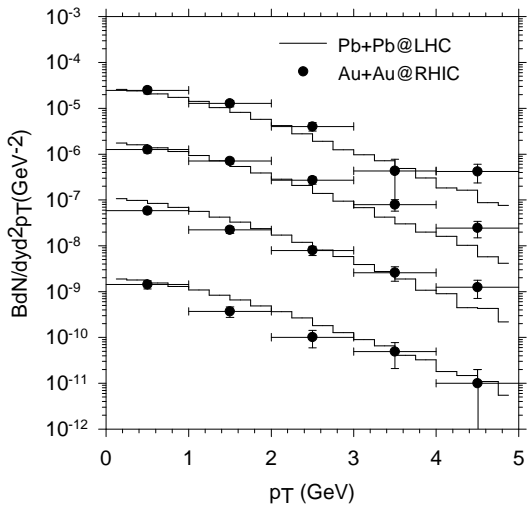}%{pbpt.eps}
\caption{ Black lines are the 'Hydro+$J/\psi$' model predictions for the $J/\psi$  $p_T$ distribution in  
0-20\%, 20-40\%, 40-60\% and 60-92\% centrality Pb+Pb collisions at LHC.
For comparison, PHENIX data for Au+Au collisions \cite{Adare:2006ns} are also shown. 20-40\% centrality collision onwards, 
the curves (and data) are progressively divided by a factor of 10.}
 \label{F7}
\end{figure}

%%%%%%%%%%%%%%%%%%%%%%%%%%%
At LHC, initial fluid temperature is high ($T_i$=421 MeV) and thermal production of charm may not be negligible. Indeed, simple estimate indicate that from RHIC to LHC, thermal charm production can increase by a factor of $\sim$ 2. Recombination of thermally produced charm and anti-charm quarks at LHC may not be negligible. Indeed, statistical hadronisation model \cite{Andronic:2007zu}
favors strong recombination effect in central Pb+Pb collisions and predict
enhancement of $J/\psi$ rather than suppression. However, we continue to neglect the recombination effect at LHC energy. As shown earlier, in the Hydro+$J/\psi$ model, existing Au+Au data do not favor recombination and the parameters controlling recombination effect at LHC energy can not be determined. The predictions for $J/\psi$ suppression given below then can be considered as an upper bound for suppression at LHC.

%%%%%%%%%%%%%%%%%%%%%%%%%%
 
In Fig.\ref{F6},   'Hydro+$J/\psi$' model prediction for
the centrality dependence of $J/\psi$ suppression in Pb+Pb collisions is shown. The melting temperature of the states and the feed-down fraction are unchanged,   
 $T_{J/\psi}=2T_c$ and $T_\chi=1.2T_c$ and F=0.3. They 
are property of the deconfined medium and do not depend on the collision energy.   In Fig.\ref{F6}, the solid line is 'Hydro+$J/\psi$' model prediction for
the survival probability in Pb+Pb collisions at LHC.  We have neglected the CNM effect here. If CNM effects are included, the suppression will increase further. For comparison, PHENIX data 
for Au+Au collisions at RHIC are also shown in Fig.\ref{F6}.  
Except for very peripheral collisions, $J/\psi$ are more suppressed at LHC energy than at RHIC energy. In central collisions, $J/\psi$'s are suppressed by a factor of $\sim$10. At LHC energy, the fluid temperature is high and melting conditions
for charmonium states are easily met than in Au+Au collisions at RHIC. Large suppression in LHC energy collisions will have experimental consequences. As stated earlier, from RHIC to LHC, particle multiplicities are expected to increase by a factor of $\sim$ 1.6. The $J/\psi$ multiplicity is also expected to
increase by a similar factor. But in central collisions from RHIC to LHC energy,  $J/\psi$ suppression
increases by a factor $\sim$ 2.5. Then effectively $J/\psi$ yield will be reduced in LHC collisions by a factor of $\sim$ 1.5. For similar statistics as in RHIC Au+Au collisions, experiments have to run for longer duration at LHC.

Lastly, in Fig.\ref{F7}, we have shown the simulated $J/\psi$ $p_T$ spectra in 0-20\%, 20-40\%, 40-60\% and 60-92\% centrality Pb+Pb collisions. For comparison, we have also shown the PHENIX data for $J/\psi$ $p_T$ spectra in Au+Au collisions \cite{Adler:2003rc,Adare:2006ns}. It appears that  $J/\psi$ $p_T$ spectra in Pb+Pb collisions at LHC is similar to that in Au+Au collisions at RHIC, may be slightly reduced in central collisions.

\section{Summary and conclusions} \label{sec5}

To summarise, in an improved version of 'Hydro+$J/\psi$' model \cite{Chaudhuri:2008qq}, we have studied  $J/\psi$  suppression and $p_T$ spectra in mid-rapidity Au+Au and  Cu+Cu collisions at RHIC. It is assumed that
in the collisions, a deconfined phase (QGP) is produced, which expands, cools, undergoes 1st order phase transition to hadronic fluid at the critical temperature ($T_c$) and then further cools to freeze-out.   The space-time evolution of the fluid is obtained by solving the 
hydrodynamic equations for ideal fluid in 2+1 dimensions \cite{QGP3}. $J/\psi$'s are assumed to be produced in initial NN collisions. They are randomly distributed according to transverse profile of the binary collision number. As the fluid evolve in time, the "free streaming" $J/\psi$'s are melted if the local fluid  temperature exceed a critical temperature $T_{J/\psi}$. Similarly, the states $\chi_c$ and $\psi\prime$ are assumed to melt above a critical temperature $T_\chi=T_{\psi\prime}=T_{\chi_c}$.  Sequential melting of the charmonium states, $\psi\prime$, $\chi$ and $J/\psi$, above critical temperatures: $T_{J/\psi}=2T_c$, $T_{\psi\prime}=T_{\chi_c}=1.2T_c$ with feed-down fraction $F=0.3$, well explain the PHENIX data  on the centrality dependence of $J/\psi$ suppression in Au+Au collisions. PHENIX data on   $J/\psi$ $p_T$ spectra in 0-20\%, 20-40\%, 40-60\% and 60-92\% centrality Au+Au collisions are also well explained. Analysis suggests that sequential melting of charmonium states, in an expanding QGP fluid, is consistent with the PHENIX data on $J/\psi$ suppression and $J/\psi$ $p_T$ distribution in Au+Au collisions at RHIC. 
Sequential melting of charmonium states on the other hand appears to be inconsistent with experimental data on $J/\psi$ production in Cu+Cu collisions.
Centrality dependence of $J/\psi$ suppression is over predicted by 20-30\%.
The $J/\psi$ $p_T$ spectra in Cu+Cu collisions are under predicted also by 
20-30\%. Within the model approximation, $J/\psi$ production mechanism in Cu+Cu collisions is at variance with the production mechanism in Au+Au collisions. 
We have also predicted for the centrality dependence of $J/\psi$ suppression and $J/\psi$ $p_T$ spectra in Pb+Pb collisions at LHC.
The model predicts that  in central Pb+Pb collisions, $J/\psi$'s are suppressed by a factor of $\sim$ 10.  The $J/\psi$ $p_T$ spectra appears to be similar to that in Au+Au collisions at RHIC.

%%%%%%%%%%%CCCCCCCCCCCCC%%%%%%%%%%%%%

\end{document}